# A Novel Tree Visualization to Guide Interactive Exploration of Multi-dimensional Topological Hierarchies

Yarden Livnat[1], Dan Maljovec[1], Attila Gyulassy[1], Dr Baptiste Mouginot[2], and Valerio Pascucci[1]

[1] SCI Institute, University of Utah, Salt Lake City, Utah, United States
[2] Nuclear Engineering, University of Wisconsin-Madison, Madison, Wisconsin, United States

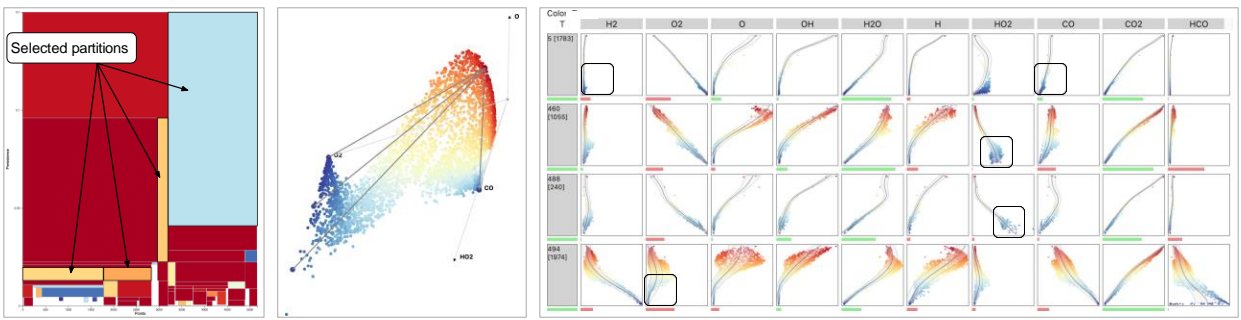

Fig. 1. A Regulus Tree provides a concise annotated overview of all Morse-Smale complex partitions for the entire persistence hierarchy. In a study of temperature as a function of chemical species, the Regulus Tree on the left encodes Child Dimension Fitness that highlights partitions with different characteristics than their parents (red/blue indicate small/large differences). Selection of the top partitions shows a single maximum and four distinct minima in the graph view (middle). Projections of the data points in the four partitions (right) demonstrates that two unique minima are due to lack of fuel (top) and lack of oxidizer (bottom), while the other two identify unmixed fuel and oxidizer due to turbulence in the combustion reaction.

**Abstract**—Understanding the response of an output variable to multi-dimensional inputs lies at the heart of many data exploration endeavours. Topology-based methods, in particular Morse theory and persistent homology, provide a useful framework for studying this relationship, as phenomena of interest often appear naturally as fundamental features. The Morse-Smale complex captures a wide range of features by partitioning the domain of a scalar function into piecewise monotonic regions, while persistent homology provides a means to study these features at different scales of simplification. Previous works demonstrated how to compute such a representation and its usefulness to gain insight into multi-dimensional data. However, exploration of the multi-scale nature of the data was limited to selecting a single simplification threshold from a plot of region count. In this paper, we present a novel tree visualization that provides a concise overview of the entire hierarchy of topological features. The structure of the tree provides initial insights in terms of the distribution, size, and stability of all partitions. We use regression analysis to fit linear models in each partition, and develop local and relative measures to further assess uniqueness and the importance of each partition, especially with respect parents/children in the feature hierarchy. The expressiveness of the tree visualization becomes apparent when we encode such measures using colors, and the layout allows an unprecedented level of control over feature selection during exploration. For instance, selecting features from multiple scales of the hierarchy enables a more nuanced exploration. Finally, we demonstrate our approach using examples from several scientific domains.

**Index Terms**—Computational Topology-based Techniques, High-dimensional Data, Data Models, Graph/Network and Tree Data. Multi-Resolution and Level of Detail Techniques.

✦

## 1 INTRODUCTION

Many phenomena in science and engineering can be described by how an output variable depends on input parameters. For example, understanding the correlation between temperature and chemical species and turbulence in a computationally model of a combustion reaction can lead to better fuel or engine designs. As another example, understanding how the measured strength of concrete varies with the ratios of its ingredients can lead to more error-tolerant mixtures. Computational models are used to study such real-world phenomena, either by conducting computer simulations or through a set of well-designed experiments. Analysis of the results can then be used to improve the models, find optimal solutions, uncover unknown relationships, and support decision-making.

The set of relationships between inputs and output can be very specialized; for any input parameter, its relationship to the output variable may be conditioned on the variation in the other parameters. Topology provides a means of studying the shape of a function; for instance, identifying how local minima and maxima are related to each other both spatially and in terms of local importance. The Morse-Smale complex, in particular, decomposes the domain into monotonic regions that enable reasoning about local trends that contribute to the formation of a local maximum or minimum. In contrast to a user-defined query or hypercube sample, the topological partitions are intrinsic to the function and underlying manifold, and are well-suited for regression analysis.

Local perturbations, artifacts of meshing, or small features can derail analysis, as it is difficult to separate phenomena from noise. Persistent homology describes topological features in terms of their life-span of the element from its birth critical point to its death in a sweep of
the range of the function. In many applications, features below a
persistence threshold are discarded as noise, a process that involves
guesstimating an appropriate value, sometimes with the help of a per-
sistence curve. In many applications, however, features appear with
varying persistence in the domain. In multi-dimensional data analy-
sis, in particular, justifying a simplification threshold is difficult as,
until now, there have not been effective visualization and exploration
techniques to understand the specific relationships between features at
different scales.

We introduce a novel visualization that is composed of a nested
space-filling tree layout to visualize the topological hierarchy whose
geometry encodes the size and persistence of topological features. We
reinterpret persistence simplification hierarchies of the Morse-Smale
complex as a merging tree of partitions, allowing an even finer gran-
ularity of feature selection than a single simplification operation and
efficient layout. Color in the cells of the tree is used to encode one of
many computed measures, such as fitness of a regression model to the
corresponding topological feature, relationships between the models
of parents and children, or any other computed attributes. Our new
visualization is deployed in an open exploration environment imple-
mented in Python and JupyterLab extensions. Linked views enable
dynamic feature selection for flexible analysis. We evaluate the utility
of the approach with use cases in combustion and nuclear energy, where
salient features are visible at a glance, that previously depended on an
exhaustive search through the simplification parameter. Specifically,
our contributions are:

- A new interpretation of persistence simplification of a Morse-
  Smale complex as a merger tree of partitions,
- A new visualization of topological hierarchies that encodes the
  size and life-span of every feature at once,
- Measures on topological features that incorporate the ancestry of
  a partition to aid and guide users in selecting the topological scale
  for analysis,
- A user interface that enables adaptive simplification, and non-
  uniform and non-consistent selection of features across multiple
  scales,
- Design of an open exploration environment to facilitate ex-
  ploratory analysis.

## 2 BACKGROUND AND RELATED WORKS

### 2.1 Topology-based analysis

Morse theory describes the topology of differentiable functions on a
manifold [22]. The critical points of a function $f$ occur where the
gradient vanishes with index equal to the number of negative eigen-
values of the Hessian. Integral lines are paths that are tangent to $\nabla f$,
do not intersect, and have lower and upper limits at critical points,
called the origin and destination, respectively. The *ascending* and
*descending manifold* of a critical point is the union of integral lines
originating, and terminating at that point, respectively. For $f$ defined
on a $d$-dimensional manifold, an index-$i$ critical point has a $d-i$ di-
mensional ascending manifold and $d$-dimensional descending manifold.
A function is Morse-Smale [27, 28] if all ascending and descending
manifolds intersect transversally, or not at all. The intersections of all
ascending and descending manifolds forms a cell complex called the
Morse-Smale complex. Each $d$-dimensional cell of the complex, called
a *partition*, is composed of points whose integral lines originate and
terminate at the same minimum-maximum pair. Figure 2 Shows the
critical points, integral lines, and cells of the Morse-Smale complex of
a 2-dimensional scalar function.

A Morse-Smale complex is simplified by repeated cancellation of
critical point pairs that differ in index by one [9]. Persistent homology
orders cancellations by increasing difference in function value [10]. In
the Morse-Smale complex, a cancellation is realized by removing a pair
of critical points and merging their ascending and descending manifolds
with their neighbors [15,16]. For $d > 2$ a 1-saddle that separates distinct
minima is called a split saddle, and a $d-1$-saddle that separates distinct
maxima is called a merge saddle. When cancelling a merge or split
saddle with an extremum, the effect on the Morse-Smale complex is

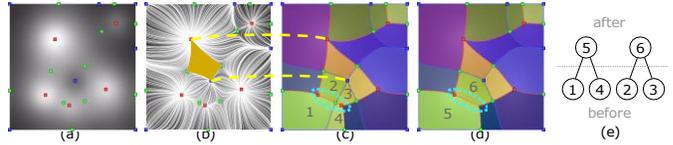

Fig. 2. The maxima (red) minima (blue) and saddles (green) of a two-
dimensional scalar function (a), are the origin and destination of integral
lines (b). A cell of the Morse-Smale complex (c) is formed by integral lines
sharing an origin and destination. The 2D patches, called a partitions
is highlighted (b,c). Cancellation of a saddle-maximum pair (circled)
merges adjacent partitions (c,d); 2,3 merge to 6, and 1,4 merge to 5.
The max-saddle cancellation is represented as these merging partition
in a Regulus Tree (e).

to merge partitions separated by the $d-1$-manifold emanating from
the saddle. As a cancellation corresponds to a *local* change in $f$,
and corresponding *local* change to the structure of the Morse-Smale
complex, independent cancellations can be organized into a *persistence
hierarchy*, a directed graph whose arrows local cancellation indicate
order dependency [3]. The subset of cancellations between extrema
and split and merge saddles turns the persistence hierarchy into a
rooted tree [30]. This tree interpretation motivates tree representation
of merging partitions, and enables the overall approach of displaying
multi-dimensional persistence hierarchies.

Morse-Smale complexes, merge trees, contour trees [6], and Reeb
graphs [25] all encode aspects of the topology of scalar functions, and
have been used successfully to define and compute features in many
application domains [2, 4]. While most methods employ persistent
homology [10] to reason about the features at multiple scales, rarely
is the merging of regions directly encoded and visualized. The branch
decomposition of contour trees comes close [24], encoding persistence
and parent-child child relationships in the merging of regions in its
structure. Persistence diagrams [8–10], and their variants, barcode
diagrams [14], and persistence landscapes [5] present the persistence
pairs and their place in the range of the function, but fail to convey the
nesting of topological features.

A standard approach for selecting an appropriate persistence value,
in the context of Morse-Smale complexes, is based on the notion of a
persistence graph, which depicts the number of extrema points with
respect to the normalized persistence value [11, 25]. The expectations
are that extrema points with low persistence are most likely due to
noise or under sampling. In addition, stable features require a large
change in the persistence value before they are removed, leading to
visible plateaus in the graph. The persistence graph, however, even
when providing clear stable thresholds, does not provide any insights
about the underlying function.

### 2.2 Exploring Parameter Spaces

An important aspect of our work is its applicability to exploration and
understanding the functional relationship between multi-dimensional
parameter spaces and their derived outputs. HyperMoVal [26] is a
system designed for validating support vector regression models against
the underlying data. The system employs linked views, local sensitivity
information, and model tuning capabilities to allow the user to see not
only where the model deviates from the data, but to refine the model
interactively until desired criteria are met.

Tuner is a software system designed to help users tune the parameters
for image segmentation and combines an automated adaptive sampling
phase with a visual exploration stage where stability and sensitivity can
be evaluated until the user guides the system to their optimal solution.
Berger et al. combine regression models and linked views to provide
users with local uncertain-aware sensitivity information that is meant
to guide users to interesting regions in their domain. Similar works
that focus on such *design steering* methodologies include the works
of Matkovic et al. [19, 20, 29] where linked views are combined with
user-guided adaptive sampling to refine the data and/or models built on
the data in areas of interest to the user.

ParaGlide [1] provides an interactive exploration of the parameter
space of multidimensional simulation models. The system enable users to define regions in the input space that represent distinct output behavior. The regions are defined manually by the user and are restricted to Cartesian product of ranges in the various input dimensions.

In contrast to these works, we use Morse-Smale complex to partition the space and persistence homology to study the space in multiple levels of details. We also fit *local* models on the data and compare and contrast them both within and across levels in the hierarchy.

## 2.3 Visual Exploration

Gerber et al. [11] were the first to use Morse-Smale approximation to visualize scalar functions defined on multi-dimensional point cloud data. They created geometric summary of a simplified Morse-Smale complex by using locally weighted regression [7] to fit inverse regression curve in each partition and use dimension reduction to embed them in a 3D display. The curves model the inverse relation from the output value to the input parameters and provide visual cues about each partition such as local and global shape, width, length, and sampling densities. Linked views showed details of the individual partitions and their one/two-dimensional relationships with respect to the output of interest. The 3D visualization is hard to interpret and, as noted by the authors, introduced phantom visual cues such as twists of the curves that did not encode real information. We follow their use of inverse regression curves when depicting details about specific partitions and for generating additional samples but we do not use the 3D view (the sampling is not part of this paper).

Maljovec et al. [18] noted that computing all the inverse regression curves in large dataset incurred high overhead and additional manipulation is required to make an aesthetically pleasing visualization where the skeleton connects at the endpoints. Instead, they employ a more abstract ball and stick style overview that required no additional computation (aside from the Morse-Smale decomposition). Their overview consist of a 2D plot of function value versus persistence value and depicted each partition as a curve between its extrema points. For data fitting, the simpler Morse-Smale regression [12, 13] strategy were computed on-demand for a specified persistence level. They also used bar charts to compare and contrast the sensitivities and goodness-of-fit of the resultant linear models. Their overview provides a consistent view of where extrema exist in the hierarchy more in line with the work presented herein, but like Gerber's work before them, still only allowed for the visualization of a single persistence level at a time. In contrast to Maljovec et al., we use the inverse regression curves and reduce the computational cost by lazily computing them only for visible partitions and only after reducing the tree. Lastly, neither work shows how a feature merges in the hierarchy and does not provide comparison *across* persistence levels.

May et al. [21] and Muhlbacher and Piringer [23] both utilize partition-based regression models that are limited to one or two dimensional axis-aligned cuts. The goals of these works are different as the idea is to build and validate regression models and understand the effects of one or two parameters on the system whereas the topology-based methods are attempting to describe the overall structure and motifs of the data such as finding similar-behaving regions in disparate areas of the domain.

## 3 REGULUS

While topological structures such as contour trees, merge trees and Morse-Smale complexes can capture features at multiple scales, they nevertheless do not describe the simplification process, nor do they provide an overview of all the simplified topologies; rather, each instance describes, and is used to explore, only one simplified topology. Conceptually, simplification consist of creating a series of progressively coarse variation of one of these topological structures. In practice, only one model is created and then transformed to describe the required simplification level. Visual exploration methods [11, 18] are also designed to visualize one simplified topology at a time. Often, phenomena of interest appear at different scales in the data, and a single simplification threshold is insufficient for analysis. The question of why should the user select a *particular* simplification level was mostly left to the user's best estimation.

In this work, we focus on the 'why' question in the context of using multi-dimensional Morse-Smale complexes to study the relationships between input parameters and the output function. Rather than develop a method to find an optimal simplification threshold, our approach is to develop a visual representation of the whole persistence space that can help guide the user exploration. The new visualization, called Regulus Tree, is based on an interpretation of the simplification process in terms of nested partitions rather than cancellation on critical points. The expressiveness of the Regulus Tree comes to light when various attributes and measures are encoded on top of it. Another consideration of our design is to empower users to define their own attributes and measures and enable on the fly modification. The Regulus Tree enables,

**Noise:** identify regions where noise is prominent

**Persistence level:** gain better understanding of the plateaus in terms of the size and stability of the partitions involves. Compare the statistical characteristics on the set of partitions for different persistence levels

**Adaptive simplification**: Select multiple persistence levels for individual features to adapt the simplification based on amount of relative rather than absolute noise, adapt to the local scale of features, and other measures of interest

**Local properties:** Compute and display local attributes of the function in different regions

**Relative measures:** Compare and contrast partitions from different locations in the function space as well as from different levels of details (persistence levels)

**Uniqueness:** Identify and study partitions that exhibit unique characteristics

**Clarity:** The Regulus Tree provide a hierarchical view of the persistence space in terms of nested partitions, which our collaborator scientists found much easier to grasp and comprehend as opposed to the technical description in terms of critical points cancellations.

In the following, we present the conceptual design, structure, and layout of the Regulus Tree as well as ways to simplify the tree itself. We describe several ways the Regulus Tree can be used for various tasks along with additional supporting views. We then introduce the notion of dynamic attributes and measures and show how they can provide unique insights and help guide the user exploration.

## 4 THE REGULUS TREE

### 4.1 The Regulus Partition Perspective

A More-Smale complex can viewed from two different perspectives. From a partition perspective, a More-Smale complex describes a tessellation of the space into monotonic partitions. From a formal perspective, it is described in terms of an intersection of ascending and descending manifolds, which generates cells including critical points, and arcs that connect them. Cancellations, although involving only a pair of critical points, may affect several partitions at once. This often poses a challenge for scientists in application domains using this approach, as the rules that govern the merging of spatial partitions are obfuscated by the simplified explanation.

Consider a single cancellation step, in which a pair of critical points is deleted as depicted in Fig. 2(c-e). From the partition perspective, the single simplification step consists of two merges of pairs of partitions. In general, and especially in multi-dimensional data, several merges can occur in each step, but each partition may participate in only one merger per step. Note that the actual simplification process stays the same and it is only our perspective that is changed. Furthermore, from the partition perspective, the partitions mergers form a *nested* hierarchy and the full simplification process forms a binary tree. The leaf nodes of the tree represent the partitions of the initial More-Smale Complex (persistence level 0), while the root represents a single partition that encompasses the whole space. To construct a Regulus Tree we first create a full More-Smale complex and then traverse the simplification list in order. For each simplification step we identify the pairs of merging partitions, create new nodes representing the merged partitions, and update the lists of internal and critical points associated with each new node (partition). Once we finish the traversal we descend down the new tree structure in a depth-first order and assign a sequential id to each node. We also reorder the data points to follow the order of the leaf nodes as described in Sect. 4.2 below.

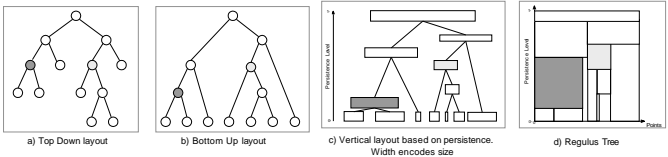

Fig. 3. Typical tree layouts position nodes based on their distance from the top (a) or bottom (b). We use vertical position to encode persistence and width to encode size (c). Regulus Tree (d): node height extends to parent to encode lifespan; horizontal layout based on enumeration of the points. Two of the nodes are shaded to illustrate the correspondence between the trees.

The notion of persistence as it applies to critical points does not directly apply to the nodes/partitions in the Regulus Tree . From the perspective of critical points, at each simplification step, one extremum is deleted while another one is retained, and no new extrema are added. The persistence of an extremum describes the value when the point is deleted. In contrast, we regard the merger of two partitions as a new partition that describes a larger region of space with more data points and different properties and characteristics. The importance of this distinction arises when we fit regression models and evaluate various measures in each partition as described in 6.3. A partition is thus associated with *two* persistence values describing its creation, original through Morse-Smale complex or through merger, and destruction, when it is merged. In the context of the Regulus Tree, we only need to save for each partition the persistence level it is created, as it is deleted at the persistence value its parent is created. The lifespan of a partition, i.e. the difference between the persistence levels of its parent and its own, provides a measure of the life-span of the partition.

### 4.2 Regulus Tree Layout

There are dozens of different ways to visualize a tree, yet conceptually all full tree layouts are based on either a top-down or a bottom-up ordering (Fig. 3). The placement of a node is based on the distance of the node from the root (top-down) or a leaf (bottom-up) in terms of the number of parent-child edges. This is true whether the layout is vertical, horizontal, or radial.

To the best of our knowledge, the Regulus Tree layout is new and unique. We describe the new layout in terms of modifying a bottom-up layout. First, we use the vertical axis to depict persistence level Fig. 3c). We then represent a tree node by a rectangle and position it vertically such that its bottom edge is aligned with the persistence level in which it is created. Because the leaves of the tree represent the base partitions, i.e. the partitions of the full More-Smale complex before any simplification, they must, by definition, have a persistence level of 0 and therefore form a single row of rectangles whose bottoms are all aligned.

When two partitions are merged, the new partition (the parent) must have a persistence level greater than that of its children and thus will be positioned vertically higher than its children. In the horizontal direction, we use the width of the rectangles to encode the number of data points and convey a measure of size. Note that if the data points were sampled uniformly, then the number of points in a partition is roughly proportional to its volume. Since, by definition, a parent contains all the points of its direct children, then the width of the parent is equal to the sum of the widths of its children. Therefore, we can position all the children of a parent sequentially in the horizontal direction without causing overlaps (Fig. 3d). Finally, we extend the top of each node to the base of its parent. The height of a node therefore encodes its lifespan since the base of the parent represents the persistence level the parent is created and the level in which the children are deleted. We note that the vertical axis of the Regulus Tree represents persistence and not function value as used in other techniques, such as the contour and merge trees, or the persistence diagram.

We can take advantage of the horizontal layout by enumerating all the data points based on the base partitions they are part of (the enumeration within a partition is not important). Using this approach, we need only two numbers per partition to indicate the range of data points that are contained in that partition. Since the parent partition contains all the data points of its children and the children are positioned sequentially, the partition's data points can also be specified via a range using two numbers. Effectively, we decoupled the data points from the hierarchical structure of the partitions and kept the memory size at $O(n + p)$, where $n$ is the number of data points and $p$ is the number of partitions in the tree.

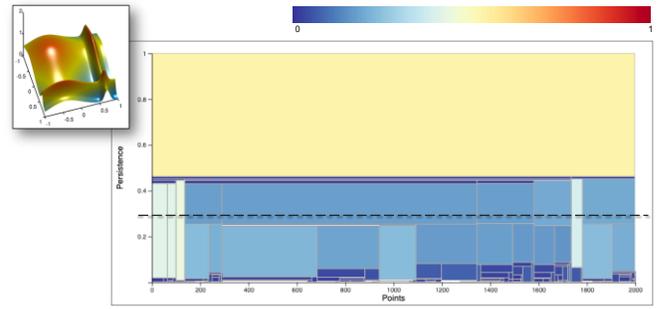

Fig. 4. A 2D scalar function and a corresponding Regulus Tree . For illustration purpose, color encodes the lifespan of a partition using the blue-yellow-red colormap shown at the top. Selecting persistence level of 0.3 amount to selecting the nodes that intersect the dashed line.

It is important to note that the above description is correct only with respect to non-critical points, which are shared between partitions. To address this, we initially assign each critical point to one of the base partitions adjacent to it. We then maintain for each partition a short list of all the critical points it's associated with but are not part of its own range of points. In the tree layout, we use the width of a node to encode only the number of points its children contain. This ensures that the parent has the correct visual width to contain all of its children. The exact number of points associated with a partition is provided in a tooltip. This does not pose any problems as the number of extra critical points is minimal.

Fig. 4 shows a Regulus Tree associated with a $2d$ scalar function, which we sampled at 2000 points and added small white noise. The horizontal axis represents enumeration of all 2000 points, while the vertical axis represents the relative persistence level in the range of 0 to 1. For illustration purposes we use color to encode the lifespan of each node, i.e the difference between the persistence value of the parent and the persistence value of the node.

The Regulus Tree is *not* a TreeMap despite the superficial similarity. A TreeMap depicts the leaves of a tree using a 2D layout that takes into consideration the tree hierarchy, and the two axes do not have individual meaning. A few variations do incorporate some information about the parents, but because the emphasis is on the leaves, the parents are depicted differently and are mainly used to convey structure. In contrast, the Regulus Tree represents the whole tree structure, the two axes have precise and different meaning and for the most part the leaves are the least important features.

### 4.3 Simplifying the Tree

Despite its compactness, the Regulus Tree can become quite large for large datasets with complex topology. In addition to pan and zoom, we can also visually simplify the tree without changing the layout by hiding nodes based on some filtering criteria. In both cases only the visualization of the tree is modified but not the tree itself. More often than not, though, we want to simplify the tree itself.

Persistence is often used to help separate between noise and real features by removing features with low persistence level, which amounts to pruning the tree at a certain threshold. From the technical perspective, persistence provides some measure of the dominance and stability of a critical point, suggesting which features should be preserved. The partition perspective of the Regulus Tree offers a different way to think

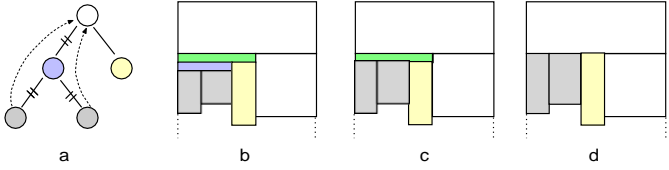

Fig. 5. Simplifying a Regulus Tree by removing intermediate nodes.

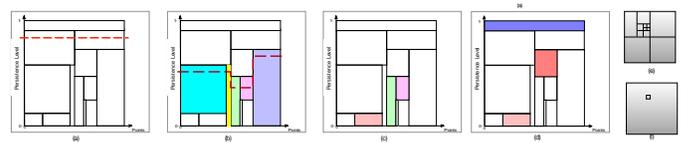

Fig. 6. Exploration strategies. a) global simplification b) adaptive simplification c) non-continuous selection d) non-consistent simplification. e-f) The rational for non-consistent simplification (see text)

of persistence. Consider the Regulus Tree depicted in Fig. 5b and the two narrow green and purple partitions, both of which have a relatively short lifespan but may actually represent a merger of two prominent features. The higher the node is located along the vertical axes, the higher the persistence levels of the two features being merged and the deeper the valley between the two mountains is (or the taller the ridge between two valleys is).

From the perspective of the Regulus Tree, such partitions are unstable and can be regarded as a relative noise within their local neighborhood. In terms of creating a simplified description of the topology, removing these partitions is akin to depicting a mountain range by only one or two mountains. We extend this notion of simplifying the topology by simplifying the Regulus Tree to also include removing small partitions (thin partitions with few points), partitions with points with values outside a range of interest, and in general filtering the function the user may wish to apply.

While removing small noise amount to pruning the tree, filtering specific nodes does not amount to removing their children. Instead we attach the children to their grandparent (Fig. 5a), which means that in general a Regulus Tree is not a binary tree. We compute a full tree simplification by traversing the tree depth-first and considering one node at a time. If many nodes are removed the new tree may end up with tall and skinny nodes. This is especially the case if we insist on keeping base partitions even if the original lifespan is too small. Our approach is to remove such base partitions and allow the tree to have a jagged edge at the bottom (Fig. 12).

We do need to ensure that the new tree represents a valid simplification of the topology. Since a node/partition is the sum of its children, it contains all of their points and thus the grandparent must by definition already contain the points of its new direct children. Another issue to consider is the lifespan of the remaining partitions. We defined the lifespan as the difference between the persistence levels of the parent and the partition. For this reason, we do not store the lifespan of a partition in the partition record; rather, we compute it on the fly based on the parent-child relationship of the node pointing to it. The lifespan of a partition is therefore relative to the tree pointing to it and a partition may have different vertical height (lifespan) after the simplification/transformation of the tree.

In general, we do not modify the original tree; rather, we create a new tree hierarchy that refers to a shared collection of partitions. In this sense, each tree is a view over the collection of partitions, similar to creating a sub-array as a view of the full array.

### 4.4 Exploration and Simplifications Strategies

In addition to providing a concise global overview, the structure of the Regulus Tree can be used to help guide the exploration and determine where to simplify,

*Global simplification*: In the context of the Regulus Tree, a persistence value maps to a single horizontal line as shown by the red dashed line in Fig. 6a. Since we do not compute level sets, it doesn't matter where the line intersects a partition, only that it does. While the persistence graph only indicates the number of active critical points for the given persistence value, the Regulus Tree provides insights about the size (width) and stability (height) of the partitions. The tree structure may, for example, reveal that a section of the persistence graph that is not a plateau is actually composed mostly of stable partitions throughout the domain except for instabilities in a small part of the domain or maybe in a number of small partitions that are likely not significant.

*Adaptive simplification*: Consider a height function depicting the geographic elevation of a valley in a mountainous area. Perturbations that might not be important in the rugged mountains may have significant importance in the flat valley area. We can apply the notion of local simplification in the context of the Regulus Tree by selecting a step like line, such a the red line in Fig. 6b. We do not alter the meaning of the simplifications. We only select a subset of the original simplifications. No new partitions are introduced.

The adaptive simplification is easy to understand in the context of the Regulus Tree but computing the boundary or interpolating a continuous function across the selected partitions might not be a trivial task as the boundary will include T-junctions. This is not an issue in the context of this work as we focus on understanding the general structure of the underlying function and identify interesting regions.

*Discrete selection*: Supporting a selection based on single persistence value is simple as it requires moving a single horizontal line up and down. The local simplification is more complex and would require interactive construction of a line with potentially multiple steps (adding and removing steps, adjusting step vertical and horizontal positions). A simpler approach is to allows the user to directly select (e.g. click on) the partitions the step line should pass through Fig. 6b.

*Non-continuous selection*: For the purpose of identifying and exploring interesting partitions, it is sufficient to select only partitions of interest (Fig. 6c) to quickly compare and contrast the properties of partitions at different persistence levels. This is by far the most often used selection method we employ in our workflows.

*Non-consistent simplification*: We can generalize the non-continuous selection by selecting partitions that overlap horizontally, that is a partition and its descendent (Fig. 6d). There are two reasons for using non-consistence simplification. One is to simply compare the properties of a partition with its parent to determine if the parent provides sufficient details. The process can be repeated up and down the tree hierarchy if a more fine tune simplification of the whole More-Smale complex is required. The second reason has to do with efficient representation. Consider a quadtree partitioning of a relatively smooth function except for one small area as shown in Fig. 6e. This space decomposition is often critical in many applications, despite the fact that 12 of the 13 partitions are very similar. On the other hand, when studying the structure of the underlying function, a more efficient representation might consist of only two nested regions as shown in Fig. 6f, which provides both a global view and local details.

## 5 VIEWS

The Regulus Tree provides an overview of the hierarchical persistence space, but it does not provide any direct view of the data points, nor does it provide direct relations between the partitions. The variable size of the nodes also makes it harder to encode multiple values for each node. In the following, we provide a short descriptions of two additional views we employ.

### 5.1 Details View

The details view (Fig. 7) depicts a set of scatterplots for a set of selected partitions where each row represents one partition and each column represents one input dimension. Each scatterplot depicts the points in the partition, where the y-axis represents the scalar function value and the x-axis represents the specific input dimension. The same y-axis range is used for all the plots across all the rows and columns. The x-axis range depends on the dimension (column) but is the same across all rows. The points are colored using the same blue-yellow-red color map and initially encode the value of the output function similar to the y-axis. Some datasets include multiple output values, only one of which is used to create the base Morse-Smale complex. The color can be used to encode any of the output variables. The partition id and the number of points in the partition (in parenthesis) are shown in the left most column.

Each plot also depicts a projection of the inverse regression curve for that partition. The semi-transparent area on both sides of the curve corresponds to one standard deviation for the corresponding input dimension.

We encode the coefficients of the linear regression models as horizontal bars under the plots. The bar in the left most column encode the intercept of the model. Green/red indicate a positive/negative coefficient respectively. The coefficients are normalized either with respect to current model or with respect to all the selected models.

### 5.2 Graph View

The graph view depicts a 2D projection of selected partitions (Fig. 9). Each partition is represented by an edge between its minimum and maximum critical points. There are many dimensional reduction methods, each with its own merits. One of a recurring complaints we receive from our collaborators is that the abstract nature of the projections often makes it very hard to comprehend and make use of. We designed the graph view to both simplify the projection and to allow the scientists to interactively explore the projections in ways that are meaningful to them. A point in multi-dimensional space is a linear combination of unit vectors each pointing along one dimension. In the Graph view we depict it as a linear combination of vectors in the 2d plane. The user can scale and rotate the vector to change its relative contributions as well as focus on specific dimensions by removing some of the vectors.

The graph view can also project the points in the selected partitions. The points colors encode the same information as in the Details view. When a partition is highlighted, the points not in that partition are rendered as small gray points. A partition is highlighted when the user hovers over the partition edge in the graph view, hover over the partition in the Regulus Tree view, or hover over the partition row in the details view. Finally, the partition's edges can be rendered by projecting their inverse curves. Although these curves are not guaranteed to end up in the appropriate critical points, they often provide a good insight about the structure of the partition.

Fig. 9 depicts three projections of the test dataset. The projection on the right demonstrates that manipulating the vectors can provide meaningful projections, in this case conveying a pseudo 3D perspective. The ability to individually manipulate each dimension proved valuable in exploring the contribution of individual and groups of dimensions. For example, by combining, subtracting and contrasting the contributions of several dimensions (same, opposite or perpendicular directions), as shown in Fig. 11.

## 6 COLORING THE TREE

The structure of the Regulus Tree provides initial insights about the More-Smale complex that describe the underlying scalar function in terms of the distribution, size, and persistence of the partitions. The expressiveness of the Regulus Tree becomes apparent when we encode additional information about the underlying scalar function. In particular, we fit linear models to the data points in each partition and compute various measures that provide insights about the local behaviour of the underlying scalar function within a partition, as well as comparison between different partitions.

### 6.1 Attributes and Measures

Each partition has several inherent attributes, such as the number of samples it contains and the persistence levels where it's created. Additional attributes, such as the min and max values of the function within the partition can be precomputed and saved. Precomputing attributes introduce several challenges. First, while some attributes are fast and cheap to compute and store, others, such as inverse regression curves, require substantial time and space, especially if precomputed for all the partitions in a large tree. Second, many partitions may not be of interest for various reasons such as if they were created due to noise in the data, have too few data points, or have a very short lifespan. Since we often filter out or hide these partitions, precomputing their attributes would be a waste of resources. Third, some attributes describe relative measures between partitions, such as between parent and child, that depends on the particular tree. For example, the lifetime of a partition is the difference between the persistence values for the creation on its parent and itself. Fourth, some attributes, such as the bandwidth used in computing reverse regression curves, depend on parameters the user may change during the exploration, and which will require to recompute them on the fly. Finally, we want to empower users to define and modify their own attributes on the fly.

Our solution is to add the notion of a measure, that is, an attribute that is defined by providing a function to compute it rather than providing its value. A measure will be lazily evaluated for a node and the value will be cached in memory, though the user can save the cached values and the measure function to a file and reloaded next time. The use of measure functions, lazy evaluation, and caching are opaque to the rest of the system, which uses them as regular maps of node id to value. Using this approach allows us to define many attributes without the computational costs, as well as add and modify attributes and measures on the fly. When visualizing a Regulus Tree, we only need to ensure the selected measure was evaluated for the visible nodes.

Often, several measure functions use similar computations, such as fitting a regression model for a given partition and then computing some derived values. We address this by defining the shared computation as a separate measure, which the other measure functions then retrieve rather than call directly.

### 6.2 Regression Models

To study the function behaviour, we employ regression analysis to fit local linear models in the various partitions. As a first step we standardize the full dataset by removing the mean and scaling to unit variance. We fit a model to each partition independently of any neighboring partitions. The main reason is that a partition can be explored in a variety of settings each leading to different sets of neighboring partitions or even none at all (Sect. 4.4). We do not want the model of a partition to change during the exploration based on indirect actions.

A linear model is expressed in terms of a set of coefficient, $\beta_i$ such as that $\tilde{y} = \sum_i^d x_i \beta_i + \beta_0$, where $\tilde{y}$ is the predicted value and $\beta_0$ is the intercept. A least square regression model is obtained by minimizing the residual sum of squares between the observed and predicted values,

$$\min_{\beta} \|X\beta - Y\|_2^2$$

To address the potential problem of multicollinearity in the linear regression, which is common in models with large number of parameters, we use Tikhonov regularization, also known as Ridge regression, which constrains the solution by imposing a penalty on the magnitude of the coefficients,

$$\min_{\beta} \|X\beta - Y\|_2^2 + \lambda \|\beta\|_2^2$$

where large $\lambda$ leads to smaller coefficients and a more robust solution to collinearity.

Our attributes and measures approach allows us to accommodate different regression models and freely switch between them during an investigation. We first define a set of model computational functions and then assign one as the current model measure, see Listing 1. Measures that depend on the regression model in a partition can fetch the 'model' attribute as shown in Listing 2. The model can be replaced at run-time, which in turn invalidates all the models that were already computed, as well as all other measures that depend on it.

Higher order regression models can also be used although they are more complex and in some sense defeat the purpose because they are not monotonic and it is difficult to interpret their coefficients. High-order model also suffers from the curse of dimensionality; a linear model requires d+1 parameters to describe an d-dimensional data but a quadratic model requires $O(d^2)$ parameters.

```
def linear_model(tree, node):
    return LinearRegresssion().fit(node.x, node.y)
def ridge_model(tree, node):
    return Ridge().fit(node.x, node.y)
tree.add_attr(ridge_model, name='model')
```

Listing 1. Using different regression models.

### 6.3 Measures

Basic measures we often use include the lifespan, minimum and maximum value, and normalize size. We also define measures to assign a unique id (encoded as different colors) to minima and maxima critical points that are shared between partitions. A shared minima/maxima measure shows the tree from a perspective of merges of minimum/maximum critical points, which in some sense is similar to a merge tree.

#### 6.3.1 Fitness

Given a linear regression model for a partition, the first question is how well the linear model actually fits the data. The More-Smale complex guarantees that the data is monotonic within a partition at persistence level 0 but it does not imply linearity. Level 0 partitions that do not have a good linear model fit imply the function was undersampled. At higher persistence levels, multiple partitions with good but different linear models might merge into a single partition with a bad fit (e.g. partitions 0 and 45 in Fig. 7). Identifying such instances can be used to determine to locally choose a persistence value lower than the merged partition.

We evaluate the fitness of a regression model using *coefficient of determination*: $R^2 = 1 - \frac{\sum(y-\tilde{y})^2}{\sum(y-\hat{y})^2}$, where $\tilde{y}$ is the predicted value and $\hat{y}$ is the mean value. The score value range between 1.0 (perfect fit) to $-\infty$. A model that always predicts the expected value of $y$ disregarding the input would have a score of 0.

**Example:** Fig. 7 (middle right) shows the Regulus Tree of the 2D function from Fig. 4. In this case, we encode the fitness score of the linear model in each partition as color after we clamped it to the range 0 (blue) to 1 (red). The details view, Fig. 7 left, shows projections of data points in several selected partitions.

In general, the higher a partition is positioned alone the vertical axis the lower the fitness score will be (less red) as each merger add more points, which by definition can only reduce the fitness. The numerical value of the current attribute is displayed via a tooltip along with additional information. Partitions 21 and 34 have very good models (0.94 and 0.96 respectively) although they are very different from each other with respect to the $x1$ dimension. This difference is reflected in their parent, partition 20, which has a lower fitness score (0.77) due to the nonlinearity the merge introduced in the $x1$ dimension.

The shallow hill (top left in the 3d view) contains over half the data points and is captured by partition 45 as a merge of four partitions with very good but very different linear models, leading to a fitness score of only (0.42). Close examination (zooming) confirms that two of the partitions merge first, followed by a merge with the third and fourth. The two intermediate partitions are not stable and have a very short lifespan. Finally, the root of the tree, partition id 0, consists of the entire domain (2000 sample points), and represent the case where we simplify the More-Smale complex all the way down to a single partition. As can

```
def fitness(tree, node):
    model = tree['model'][node]
    return model.score(node.x, node.y)
def parent_fitness(tree, node):
    parent_model = tree['model'][node.parent]
    return parent_model.score(node.x, node.y)
def child_fitness(tree, node):
    model = tree['model'][node]
    return model.score(node.parent.x, node.parent.y)
```

Listing 2. Fitness score (not suitable for derived trees)

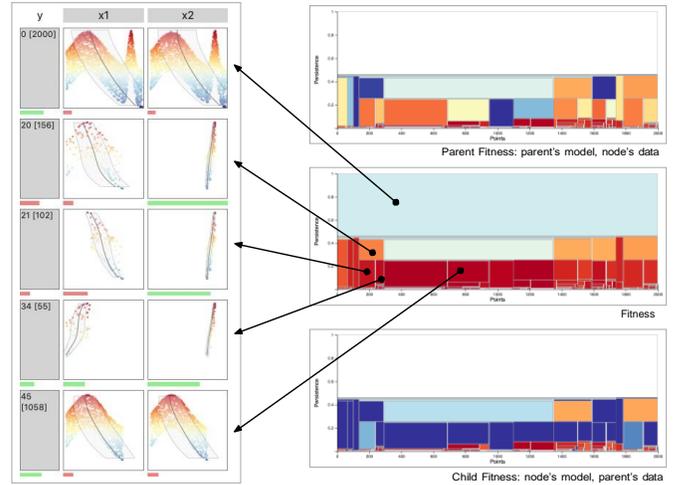

Fig. 7. Left: details view showing points in selected partitions. Right: Three views of the Regulus Tree, each showing separate fitness measure (blue = 0, red = 1)

be expected, the fitness score is only 0.38 since there is no good global linear model for the data.

**Parent and Child Fitness** Partitions 20, 21 and 34 in the above example, highlight the case where a merger of two partitions, both with very good linear models, can lead to a partition with a much worse fitness score. In this case, we should avoid simplifying further at least locally. The different situation can arise where the parent model is very similar one of the children but not the other one. Fig. 8 depict a potential merger between two partitions for a 1d scalar function, where both the children and the parent have good linear models. If we rely only on the fitness score of the parent, we may conclude that the parent represents a good simplification choice, a decision we may not take if we examine at the actual data. This scenario can occur, for example, when one partition contains a lot more data points then the second partition. In some applications, this can be addressed by giving different weights to the points in the two partition. In the context of this work, we specifically want to identify and flag situations like this as they are likely pointing region where the scalar function have unique characteristics. Furthermore, we would like to be able to detect these cases directly in the Regulus Tree without the need to examine the data points.

To address this issue we introduce two *relative* fitness measures as shown in Listing 2. The *Parent Fitness* is the fitness score of the parent's model with respect to the partition data. Conversely, the *Child Fitness* is the fitness score of the partition model with respect the parent's data. Referring back to Fig. 7, the parent fitness is shown in the top right tree and the child fitness in the bottom right tree. The parent fitness indicates that the model of partition 20 moderately fits the data in partitions 21 and 34 (0.79 and 0.65 respectively). The child fitness scores clearly show that the models for partitions 21 and 34 do not fit well with their parent's data (0.2, -1.8 respectively), strongly implying that this simplification step should not be used.

**Example part II:** Using the combination of fitness and child fitness scores, we can see that a persistence level of 0.2 provides a very good simplification value. Fig. 9 shows a top down projection of the full

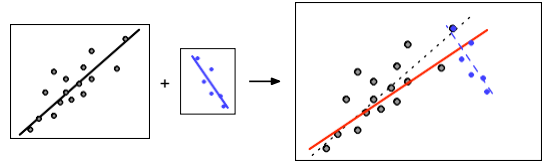

Fig. 8. A merge of two partitions with good but different models can lead to a partition with a model that is similar to only one of them.

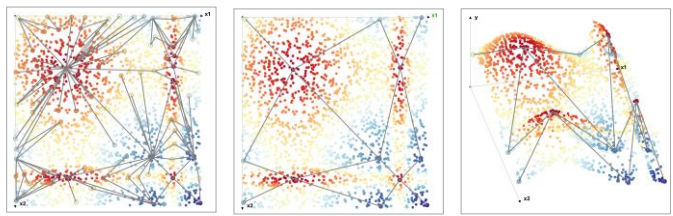

Fig. 9. Left: Graph view of the full More-Smale complex and the data points as viewed from above. Middle: Simplification for persistence level = 0.2 Right: Adding the y-axis and slightly rotating the x1- and x2 axis generates a pseudo 3D projection.

More-Smale complex and both top down and side view projections of the simplified More-Smale complex . We designed our visual exploration environment specifically to cater to this kind of workflow where multiple measures need to be considered at the same time. The user can visualize multiple Regulus Tree instances, each encoding different measures, or simply define a new measure that returns a value based on evaluation of the the fitness, child fitness, and parent fitness in the partition.

**Reference Model Fitness:** The parent and child fitness measures play an important role in our workflow despite being limited to only the local neighborhood of a node. As a side note, we do not define a fitness measure between siblings because in the case of derived trees (see Sect. 4.3 a partition can have more than two children. The advantage of the local nature of the parent and child fitness measures is that we can depict each measure over all the nodes in the tree at once. There are cases, however, where we want to compare and contrast models for more distanced ancestors and even between partitions in different parts of the tree (recall that the tree is organized with respect to the list of merges generated using persistence homology not based on local geometry). Depicting many independent global comparisons at once is not possible. Instead, we define a *reference model fitness* measure that applies a reference model to the data in each partition. When the user hovers over the tree we set the reference model to the model of the node under the mouse and reset the cache to cause the values to be recomputed.

**Dimension Fitness:** Regression methods fit linear regression models by minimizing cost functions that take into consideration all the dimensions of the data. In that sense the cost (error) is spread over all the dimensions. Within the context of this work, we are trying to identify unique partitions, which means we aim to find maximum discrepancies between partitions and in particular with respect to individual dimensions.

Our *dimension fitness* approach is to compute a vector of regression models for each partition, one per dimension, instead of a single model. Scoring in this case means applying each model in the vector to the data, resulting in a vector of scores instead of one value. Given two partitions, we apply and score the model vector of one partition to both data sets and finally compute the cosine similarity between the two score vectors (Listing 3).

```
def relative_dim(tree, models, data):
    return [models[i].score(data.x[i], data.y)
            for i in range(len(models))]
def child_dim_fitness(tree, node):
    models = tree['dim_models'][node]
    return cosine_similarity(
        tree['relative_dim'][models, node],
        tree['relative_dim'][models, node.parent])
```

Listing 3. Dimension Fitness

## 7 CHAINED ATTRIBUTES

Since several trees can point to the same collection of partitions, we would have preferred to store attributes in their respective partition

```
def relative-fitness(tree, has_model, has_pts):
    return tree['model'][has_model]
            .score(has_pts.x, has_pts.y)
def parent-fitness(tree, node):
    return tree['relative_fitness'][node.parent, node]
def child-fitness(tree, node):
    return tree['relative_fitness'][node, node.parent]
```

Listing 4. Chained attributes. Parent/child relation depends on the current tree structure

to avoid recomputing them for each tree. On the other hand, relative measures are defined with respect to a tree structure and thus should be stored at the nodes. This means that when looking for an attribute or measure for a partition, we need to know where to search for it, which undermine the notion of separation of concerns. Derived trees, such as when one tree is a simplification of another, add further complication. The new derived tree will most likely have a similar structure to the original tree but will be composed of new nodes. We would rather not have to copy or recompute and instead reuse those relative attributes the are valid for the new tree.

Our solution is to chain attributes by maintaining a pointer to the attributes of the parent tree. When the value of an attribute is not found in the new tree, we first consult the parent's attributes before computing the value. To avoid recomputing relative measures, we stored such measures using a key consisting of the ids of both partitions. We can then reuse a previously computed value if the same parent-child pair exist in the original tree. To support this we set the node id to its partition id so that keys stays the same in derived trees. Listing 4 show how we redefine parent and child fitness in terms of a relative measure (compare to Listing 2)

## 8 USE CASES

### 8.1 Combustion

In this example we look at sample data extracted from a time dependent jet simulations of turbulent CO/H2-air flame, where each sample point consists of chemical composition and temperature [17]. The data includes extinction and reignition phenomena where several chemical components form and evolve during the combustion reaction and in turn effect the amount of heat released. In this analysis we explore the temperature in relation to the chemical composition.

The data consist of 5172 samples with 10 chemical species. Fig. 10, depicts fitness for a Regulus Tree after filtering out partitions with less then 100 data points. The root of the tree demonstrates that a single linear model describe the whole data with an exceptional score of 0.998. We fit good models in most other partitions but they are not necessarily similar to each other. A zoom-in to persistence range [0, 0.15] (middle) reveals a large partition containing 30% of data points all of which lie outside of the active combustion zone and should have been removed. This fact wasn't noticed in previous works.

Using child and parent fitness doesn't reveal much but once we switch to child dimension fitness the tree comes to life, Fig. 1. The figure also shows the data points of the (four) top-most partitions that are different than their parents. Partition 494 (right most in the tree, bottom in the grid) is distinctly different from the other partitions. The three other partitions differ mainly with respect of HO2, though the points in the first and third partitions (5 and 488) are mostly in the lower value. Fig. 11 shows the critical points graph. Note that the edges for the first three partitions overlap in (b) but are clearly separated in (c).

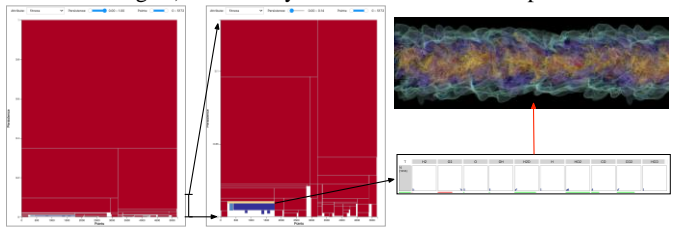

Fig. 10. Combustion: Initial tree (left) shows that a single linear model describes all the data points(score = 0.998). Right: a zoom to persistence range [0:0.15] reveal a previously unknown partition containing large number ( 30%) of sample points that are outside of the active combustion zone and should have been removed (top).

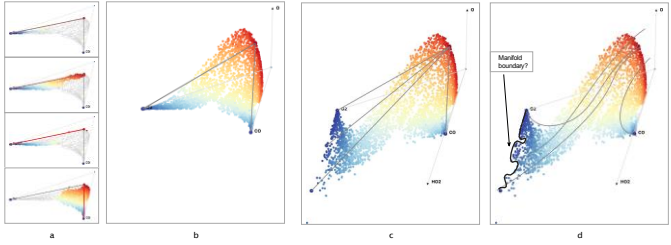

Fig. 11. Graph views of the combustion data. a) Highlighting the points of each partition. b) Without the HO2 only the two expected minima are apparent. c) The three minima separates along the HO2 direction. d) Regression curves follow the shape of the manifold.

The four partitions share a single maxima but feature four distinct minima. The minima in partition 494 captures a situation where fuel (H2 and CO) is available but the lack of oxidizer (O2) prevents a chemical reaction. The situation is reversed in partition 5 where oxidizer (O2) is available but the lack of fuel prevents a reaction. In partitions 460 and 488, the mixing of fuel and oxidizer is highly turbulent and blows the flames out, resulting in large amount of HO2. The clear separation between these two minima could be due to undersampling or possible the boundary of the manifold.

### 8.2 Nuclear Fuel Cycle Simulations

Nuclear fuel cycle analysis focuses on modeling the nuclear industry and ecosystem at a macroscopic level. This example studies scenarios for transitioning from one technology, Light Water Reactors (LWR) to a newer Sodium-cooled Fast breeder Reactor (SFR) technology. LWR reactors can use either enriched uranium (UOX, Uranium Oxyde) or a mixture of Uranium and Plutonium (MOX, Mixed Oxyde) as fuel but produce Minor Actinides waste. Minor Actinides have a long lifetime and high activities, which make such wastes difficult to deal with. In contrast, SFR reactors mainly use a mix of Natural Uranium and Plutonium as a fuel (MOX). The SFR reactors have the ability to breed Plutonium from the Uranium and energy production is based on the fission of the Plutonium. This breeding capability allows the fuel to stay longer in the fuel, reducing the amount of Minor Actinides ultimately present in the waste. Some combination of fuel and SFR reactor configuration allows to breed more Plutonium than it will burn . A sufficiently large number of SFR reactors (used in 'breeder' configuration) can thus be self sustained.

The study consists of 3300 simulation runs, using four input parameters (breeding ratio, start year of LWR fuel reprocessing, first year an SFR can be deploy and a bias measure). The aim here is to find a deployment schedule that transitions from an ecosystem consisting of LWRs to one with only SFR reactors, while minimizing some objective function. To this end we computed several objective functions at the end of each simulation including: the ratio of total power generated by LWRs reactors to the total energy generated over the simulation's

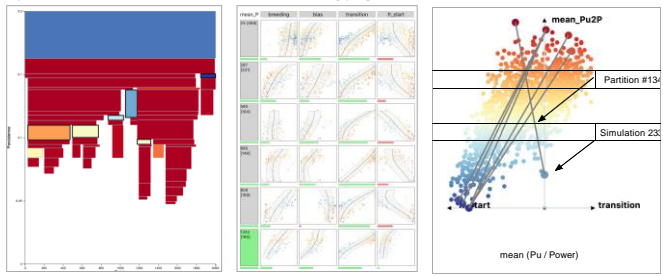

Fig. 12. Transition scenario: Partitions selected based on dimension fitness (left). Partition 1342 exhibits unique behaviour (mid and right) though its minimum (simulation 2330) is not the lowest (right).

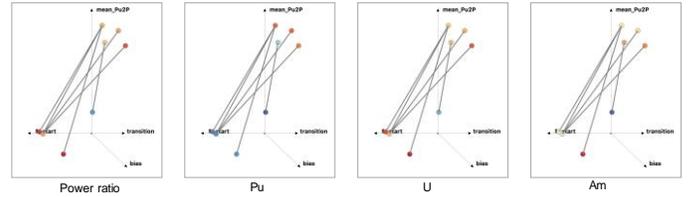

Fig. 13. Encoding other output variables as color demonstrates the advantages of simulation 2330 which exhibits low power ration and consistently generates low volumes of excess radio active materials.

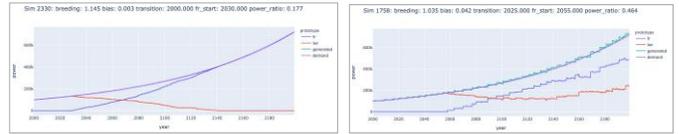

Fig. 14. Simulation 2330 exhibits an excellent and smooth transition (left), while simulation 1758 does not lead to a transition (right).

200 years span, the mean ratio of plutonium to generated power, and amount of nuclear waste such as Plutonium, Uranium and Americium. Of the 3300 simulation only 2007 led to a complete transition within the first 120 years). In the following we looked at the mean ration of Plutonium to power objective function. Fig. 12 show a reduced Regulus Tree (remove partitions with less than 100 simulations or a lifespan less then 0.001) depicting child dimension fitness. Several partitions that stand out are also shown. The graph view on the right shows that partition 1342 exhibits a unique behaviour, although its minima (simulation run 2330) doesn't have the lower value. Fig. 13 shows that adding 'bias' in the graph view affects only one of the minima. However, when we use the same partitions and graph, but change the colors to encode other output values, we can see that simulation 2330 is unique in that it consistently exhibits low values (blues) for power ration and the amount of the nuclear waste generated. Fig. 14 shows the deployment schedule of simulation 2330 (top) and an example of a simulation that doesn't lead to a complete transition.

## 9 CONCLUSIONS

The Regulus Tree addresses the important, though often neglected, 'why' question by proposing a new perspective of the topology simplification process. The Regulus Tree visualization offers both a concise broad view of the simplification landscape and a guide for an interactive visual exploration of the underlying scalar function. We describe the Regulus Tree in the context of Morse-Smale complexes, but the partition's perspective and the tree are equally applicable to Morse complexes. Some of the measures as well as the inverse regression curves are not directly applicable and one will have to use high order regression models.

The Regulus Tree has several limitations. First, it does not preserve spatial locality or even adjacency relation between partitions. Mapping spatial locality is a complex issue for multi-dimensional data in general. Adjacency information can be retrieved from the Morse-Smale complexes, though how to depict it is not clear and is especially problematic in a setting with many levels of details. Second, Morse-Smale partitions can have complex twisting shapes that are not captured directly in the tree structure and the tree does not address the notion of topological holes.

We have begun exploring methods for using the inverse regression curves to facilitate adaptive sampling, both for validation purposes and for improving spatial resolutions in areas that are undersampled. We are also looking at using t-SNE and other dimensional reduction and clustering techniques to analyze the linear models and provide additional measures for identifying and highlighting potential unique partitions.

## REFERENCES

[1] S. Bergner, M. Sedlmair, T. Moller, S. N. Abdolyousefi, and A. Saad. Paraglide: Interactive parameter space partitioning for computer simulations. *IEEE Transactions on Visualization and Computer Graphics*, 19(9):1499–1512, Sep. 2013. doi: 10.1109/TVCG.2013.61

[2] H. Bhatia, A. G. Gyulassy, V. Lordi, J. E. Pask, V. Pascucci, and P. Bremer. Topoms: Comprehensive topological exploration for molecular and condensed-matter systems. *Journal of Computational Chemistry*, 0(0), 2018. doi: 10.1002/jcc.25181

[3] P. . Bremer, B. Hamann, H. Edelsbrunner, and V. Pascucci. A topological hierarchy for functions on triangulated surfaces. *IEEE Transactions on Visualization and Computer Graphics*, 10(4):385–396, July 2004. doi: 10.1109/TVCG.2004.3

[4] P.-T. Bremer, G. Weber, V. Pascucci, M. Day, and J. Bell. Analyzing and tracking burning structures in lean premixed hydrogen flames. *IEEE Transactions on Visualization and Computer Graphics*, 16(2):248–260, 2010.

[5] P. Bubenik. Statistical topological data analysis using persistence landscapes, 2012.

[6] H. Carr, J. Snoeyink, and U. Axen. Computing contour trees in all dimensions. *Computational Geometry*, 24(2):75–94, 2003.

[7] W. S. Cleveland and S. J. Devlin. Locally weighted regression: An approach to regression analysis by local fitting. *Journal of the American Statistical Association*, 83(403):596–610, 1988. doi: 10.1080/01621459.1988.10478639

[8] D. Cohen-Steiner, H. Edelsbrunner, and J. Harer. Stability of persistence diagrams. *Discrete & Computational Geometry*, 37(1):103–120, Jan 2007. doi: 10.1007/s00454-006-1276-5

[9] H. Edelsbrunner, J. Harer, and Zomorodian. Hierarchical morse smale complexes for piecewise linear 2-manifolds. *Discrete and Computational Geometry*, 30:87–107, 07 2003. doi: 10.1007/s00454-003-2926-5

[10] H. Edelsbrunner, D. Letscher, and A. Zomorodian. Topological persistence and simplification. *Discrete Comput Geom*, 28:511–533, 2002.

[11] S. Gerber, P. Bremer, V. Pascucci, and R. Whitaker. Visual exploration of high dimensional scalar functions. *IEEE Transactions on Visualization and Computer Graphics*, 16(6):1271–1280, Nov 2010. doi: 10.1109/TVCG.2010.213

[12] S. Gerber and K. Potter. Data analysis with the morse-smale complex: The msr package for r. *Journal of Statistical Software*, 50(2):1–22, 7 2012.

[13] S. Gerber, O. Rübel, P.-T. Bremer, V. Pascucci, and R. T. Whitaker. Morse-smale regression. Manuscript, 2011.

[14] R. Ghrist. Barcodes: The persistent topology of data. *BULLETIN (New Series) OF THE AMERICAN MATHEMATICAL SOCIETY*, 45, 02 2008. doi: 10.1090/S0273-0979-07-01191-3

[15] A. Gyulassy, N. Kotava, M. Kim, C. D. Hansen, H. Hagen, and V. Pascucci. Direct feature visualization using morse-smale complexes. *IEEE Transactions on Visualization and Computer Graphics*, 18(9):1549–1562, 2012.

[16] A. Gyulassy, V. Natarajan, V. Pascucci, P.-T. Bremer, and B. Hamann. Topology-based simplification for feature extraction from 3D scalar fields. *IEEE Transactions on Computer Graphics and Visualization*, 12(4):474–484, 2006.

[17] E. R. Hawkes, R. Sankaran, J. C. Sutherland, and J. H. Chen. Scalar mixing in direct numerical simulations of temporally evolving plane jet flames with skeletal co/h2 kinetics. *Proceedings of the combustion institute*, 31(1):1633–1640, 2007.

[18] D. Maljovec, B. Wang, P. Rosen, A. Alfonsi, G. Pastore, C. Rabiti, and V. Pascucci. Rethinking sensitivity analysis of nuclear simulations with topology. In *2016 IEEE Pacific Visualization Symposium (PacificVis)*, pp. 64–71. IEEE, 2016.

[19] K. Matković, D. Gračanin, and H. Hauser. Visual analytics for simulation ensembles. In *Proceedings of the 2018 Winter Simulation Conference*, WSC '18, p. 321–335. IEEE Press, 2018.

[20] K. Matković, D. Gračanin, R. Splechtna, M. Jelović, B. Stehno, H. Hauser, and W. Purgathofer. Visual analytics for complex engineering systems: Hybrid visual steering of simulation ensembles. *IEEE Transactions on Visualization and Computer Graphics*, 20(12):1803–1812, Dec 2014. doi: 10.1109/TVCG.2014.2346744

[21] T. May, A. Bannach, J. Davey, T. Ruppert, and J. Kohlhammer. Guiding feature subset selection with an interactive visualization. In *IEEE VAST*, pp. 111–120, 2011. doi: 10.1109/VAST.2011.6102448

[22] J. Milnor. *Morse Theory*. Princeton University Press, 1963.

[23] T. Muhlbacher and H. Piringer. A partition-based framework for building and validating regression models. *IEEE Transactions on Visualization and Computer Graphics*, 19(12):1962–1971, 2013.

[24] V. Pascucci, K. Cole-McLaughlin, and G. Scorzelli. The toporrery: computation and presentation of multi-resolution topology. In T. Möller, B. Hamann, and R. D. Russell, eds., *Mathematical Foundations of Scientific Visualization, Computer Graphics, and Massive Data Exploration*, pp. 19–40. Springer Berlin Heidelberg, Berlin, Heidelberg, 2009. doi: 10.1007/b106657 2

[25] V. Pascucci, G. Scorzelli, P.-T. Bremer, and A. Mascarenhas. Robust on-line computation of reeb graphs: simplicity and speed. In *ACM SIGGRAPH 2007 papers*, pp. 58–es. 2007.

[26] H. Piringer, W. Berger, and J. Krasser. Hypermoval: Interactive visual validation of regression models for real-time simulation. In *Proceedings of the 12th Eurographics / IEEE - VGTC Conference on Visualization*, EuroVis'10, pp. 983–992. Eurographics Association, 2010.

[27] S. Smale. Generalized Poincaré's conjecture in dimensions greater than four. *Ann. of Math.*, 74:391–406, 1961.

[28] S. Smale. On gradient dynamical systems. *Ann. of Math.*, 74:199–206, 1961.

[29] R. Splechtna, K. Matković, D. Gračanin, M. Jelović, and H. Hauser. Interactive visual steering of hierarchical simulation ensembles. In *2015 IEEE Conference on Visual Analytics Science and Technology (VAST)*, pp. 89–96, Oct 2015. doi: 10.1109/VAST.2015.7347635

[30] J. Tierny and V. Pascucci. Generalized topological simplification of scalar fields on surfaces. *IEEE Transactions on Visualization and Computer Graphics*, 18(12):2005–2013, 2012.